\documentclass[showpacs,prl,onecolumn,aps,superscriptaddress,preprintnumbers,nofootinbib]{revtex4}
\usepackage[T1]{fontenc}
\usepackage[latin9]{inputenc}
\usepackage{amsmath,amssymb}
\usepackage{epsfig}
\usepackage{graphicx}
\usepackage{mathrsfs}
\usepackage{amsmath}
\usepackage{amsfonts}
\usepackage{epstopdf}
\usepackage{color}
\def\slashchar#1{\setbox0=\hbox{$#1$}     		
   \dimen0=\wd0                                 	
   \setbox1=\hbox{/} \dimen1=\wd1               	
   \ifdim\dimen0>\dimen1                        	
      \rlap{\hbox to \dimen0{\hfil/\hfil}}      	
      #1                                        	
   \else                                        	
      \rlap{\hbox to \dimen1{\hfil$#1$\hfil}}   	
      /                                         	
   \fi}

\newcommand{\beq}{\begin{equation}}
\newcommand{\eeq}{\end{equation}}
\newcommand{\bea}{\begin{eqnarray}}
\newcommand{\eea}{\end{eqnarray}}
\newcommand{\ba}{\begin{array}}
\newcommand{\ea}{\end{array}}

\def\eq#1{{Eq.~(\ref{#1})}}
\def\fig#1{{Fig.~\ref{#1}}}
\newcommand{\bas}{\bar{\alpha}_S}

\newcommand{\nn}{\nonumber}

\newcommand{\Lb}{\left(}
\newcommand{\Rb}{\right)}
\newcommand{\h}{\frac{1}{2}}

\newcommand{\pom}{I\!\!P}
\newcommand{\Y}{\tilde{Y}}

\begin{document}

\newcommand{\pp}{\partial}
\newcommand{\intl}{\int\limits}

\title{Can $1/N_c$   corrections be treated in the Pomeron calculus?}

\author{Eugene Levin}
\email{leving@tauex.tau.ac.il}
\affiliation{Department of Particle Physics, Tel Aviv University, Tel Aviv 69978, Israel}

\date{\today}


\begin{abstract}
The main goal of the paper is to show that we can treat the $1/N_c$ QCD corrections in the Pomeron calculus.  
We develop the one dimensional model which is a simplification of the QCD approach that includes $\pom \to 2 \pom$, $2 \pom \to \pom$ and $ 2 \pom \to 2 \pom$ vertices and
gives the description of the high energy interaction, both in the framework of the parton cascade and in the Pomeron calculus. In this model we show that
  the scattering amplitude can be written as the sum of Green's function of $n$-Pomeron exchanges $G_{n \pom} \propto e^{ \omega_n \Y}$ with $\omega_n = \kappa\,n^2$ at $\kappa \ll 1$.  This means that choosing $\kappa = 1/N^2_c$ we can reproduce the intercepts of QCD in $1/N_c$ order.  The scattering amplitude is an asymptotic series that cannot be sum using Borel approach. We found a general way of summing such series.   In addition to the positive  eigenvalues we found the set of negative   eigenvalues which corresponds to the partonic description of the scattering amplitude. 
 Using Abramowsky, Gribov and Kancheli  cutting rules we found the multiplicity distributions of the produced dipoles as well as their entropy $S_E$.

 \end{abstract}
\maketitle
\vspace{-0.5cm}
\tableofcontents


\section{Introduction}

The question in the title of this paper arises from  the rapid increase of  theGreen's function of the $n$-BFKL Pomeron exchanges  ($G_{n \pom}$)\cite{BFKL,LIP,LIREV} which intercept is proportional to $\frac{\bas}{N^2_c} n^2$ ($\Delta_{n\pom} \propto \frac{\bas}{N^2_c} n^2$)  as it has been shown using the BKP\footnote{BKP stands for Bartels, 
Kwiecinski and Praszalowicz.} equation in Refs.\cite{LRS,BART0,BARTU,LLS,LLR}\footnote{See also the short and pessimistic review of the problem in Ref.\cite{LENC}.}. Such an increase leads to the scattering amplitude in the BFKL Pomeron calculus being the asymptotic series of  Green's functions that cannot be summed using the Borel approach \cite{BORSUM}. 

The goal of this paper is to learn how to sum such asymptotic series. This problem is very old one and in the Pomeron calculus  it
goes under the slang name  of many particle Regge poles      \cite{DISASTER}.
For long time most experts me including believed that the asymptotic series with $\Delta_{n \pom} > n \Delta_{\pom}$ cannot be summed in the framework of the Pomeron calculus and we have to find a different approach.
 However, recently in Ref.\cite{UTMM}  it  has been demonstrated that  it is not correct. In this paper it is shown that the one dimensional model which satisfies both $t$ and $s$ channel unitarity\cite{MUSA,BIT,RS,KLremark2,UNRFT,UTM} leads to $\Delta_{n \pom} = \exp\Lb \gamma \,n\Rb - 1$ where $\gamma$ is the scattering amplitude of two dipoles at low energy. Nevertheless,  the way to sum the asymptotic series for the scattering amplitude has been found. The method proposed in Ref.\cite{UTMM}, is rather general: the analytic continuation of every $G_{n\pom}$ to negative $n$. After such continuation each $G_{n \pom} \propto \exp\Lb\Lb  e^{- n\,\gamma}\,-\,1\Rb Y\Rb$ where $Y$ is rapidity. The scattering amplitude turns to the sum of the exponentially small contributions at large $Y$. On the  other hand it is clear that for the intercepts proportional to $n^2$ this method does not work. The second feature of the model: it has infinite number of the pomeron vertices. It is not clear how this can be seen in QCD. 
 
 In this paper we consider the one dimensional model with restricted set of Pomeron vertices : $\pom \to 2 \pom$, $2 \pom \to \pom$ and $ 2 \pom \to 2 \pom$.  The parton cascade for such interactions has been developed in QCD in Ref.\cite{LELU1}.  It is instructive to note that the Pomeron calculus with  $\pom \to 2 \pom$ and  $2 \pom \to \pom$ corresponds to the Braun lagrangian in QCD\cite{BRAUN}   which reproduces the nonlinear Balitsky-Kovchegov (BK) equation  in arbitrary reference frame\footnote{ As it was shown in Ref.\cite{LELU2}(see also Ref.\cite{UTM}) the BFKL cascade leads to the BK equation only in the l.r.f. with a nucleus  being  at the rest.}.
 We show that this cascade  generates  $1/N^2_c$ corrections and their treatment is closely related to another
   unsolved theoretical problem \cite{BFKL,LIP,LIREV,LELU1,KOLEB,LIFT,GLR,GLR1,MUQI,MUDI,Salam,NAPE,BART,BKP,MV,MUSA,KOLE,BRN,BRAUN,BK,KOLU,JIMWLK1,JIMWLK2,JIMWLK3,JIMWLK4,JIMWLK5,JIMWLK6,JIMWLK7,JIMWLK8,AKLL,KOLU1,KOLUD,BA05,SMITH,KLW,KLLL1,KLLL2,LEN}  in QCD: summation of the BFKL Pomeron loops\footnote{ The abbreviation BFKL Pomeron  stands  for Balitsky, Fadin, Kuraev and Lipatov Pomeron.}. Indeed, in Refs.\cite{LRS,BART0,BARTU,LLS,LLR} it has been shown
    that the large $1/N^2_c$  corrections to the intercepts opf $G_{n \pom}$   can be treated in the BFKL Pomeron calculus if we introduce the vertices of interaction of four Pomerons ($2 \pom  \to  2 \pom $).

     \begin{figure}[ht]
    \centering
    \begin{tabular}{c c}
  \leavevmode
      \includegraphics[width=8.9cm]{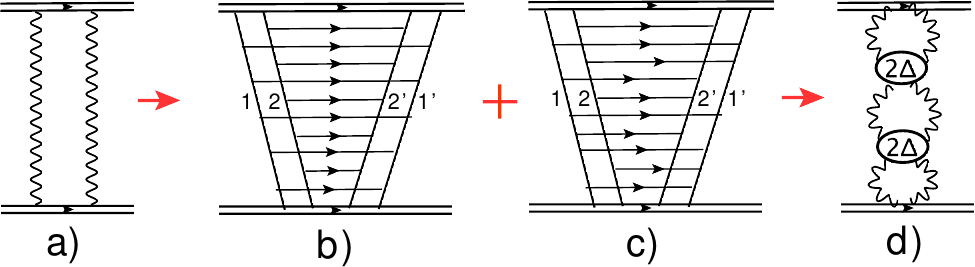}& \includegraphics[width=2.8cm]{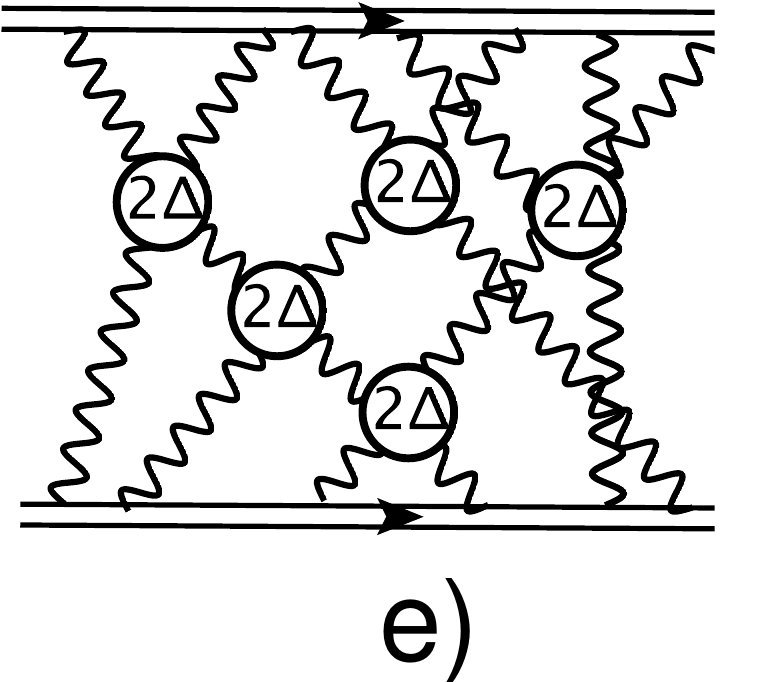}\\
      \end{tabular}   
         \caption{ Summing the switch diagrams for the exchange of two BFKL Pomerons.
      \fig{2pom}-a: the exchange of two BFKL Pomerons. \fig{2pom}-b: the exchange of two BFKL Pomeron in the two dimensional model.  \fig{2pom}-c: the switch diagram for two ladder exchange in which the produced dipoles from the Pomeron 1 are absorbed by the Pomeron 2' and vise versa.  \fig{2pom}-d: the sum of all diagrams in the two dimensional model that contribute in the Green's  function of the exchange of two Pomerons in t-channel.  \fig{2pom}-e: the Green's function of the exchange of n$n$ Pomerons.The wavy lines denote the exchange of one Pomeron with the Green's function $\exp\Lb \Delta Y\Rb$. The $\pom + \pom \to \pom + \pom$ vertex is denoted by blob and equals to $2 \Delta \kappa$ in QCD-type model.}
\label{2pom}
   \end{figure}
 In this introduction we want  to emphasize that the $1/N^2_c$ corrections have a general origin which has been discussed in seventies\cite{DISASTER}. We illustrate this origin  considering the two Pomeron exchange which is shown in \fig{2pom}-a. This Pomeron diagram  corresponds to the Feynman diagram of \fig{2pom}-b in which all dipoles emitted by the ladder 1 are absorbed by ladder 1', and all  dipoles produced  by ladder 2 are absorbed by ladder 2'. These diagrams lead to the Green's function of the exchange of two Pomerons: $G_{2 \pom} =G^2_{\pom}\Lb Y\Rb $.  However, the diagrams in which  the dipole,  emitted from ladder 1, will be absorbed by ladder  2'  are not small. We are going to call these diagrams  ``switch diagrams" , using the terminology suggested in Refs.\cite{LRS,BART0,BARTU,LLS,LLR}.  Indeed, after first exchange ladder 1 and 2' will give the Pomeron exchange, leading to the diagram of \fig{2pom}-d. Since at given rapidity we have two `switch diagrams: dipole emitted between ladders 1 and 2' and the dipole emitted between ladders 2 and 1' we can  obtain the vertex $\pom + \pom \to \pom + \pom$ , which is equal to $ 2 \Delta$. In QCD this vertex has $1/N^2_c$ suppression and we denote it as $2 \kappa\,\Delta$ with $\kappa \sim 1/N^2_c$.

The diagrams of \fig{2pom}-d can be summed  in $\omega $-representation:

\beq \label{POM3}
G_{2 \pom}\Lb \omega \Rb\,\,=\,\,\sum_{k=0}^\infty \frac{1}{\omega\,-\,2\,\kappa\,\Delta} \Lb \frac{2\,\Delta\,\kappa}{\omega\,-\,2\,\kappa\,\Delta}\Rb^k\,=\,\frac{1}{\omega\,-\,2\,\Delta - 2 \kappa \Delta}
\eeq
Where $1/\Lb \omega - 2 \Delta\Rb$ is the contribution of the exchange of two Pomerons (see \fig{2pom}-b).  Coming back to $Y$ representation one can see that $G_{2 \pom}\Lb Y\Rb\,\,=\,\,\exp\Lb 2\Lb 1\,+\,\,\kappa\Rb \,Y\Rb$ instead of $G_{2 \pom}\Lb Y\Rb\,\,=\,\,\exp\Lb 2\,\Delta\,Y\Rb$  which is expected for two Pomeron exchange. Summing diagrams of \fig{2pom}-e one can see that the intercept of $G_{n \pom}$ is equal to  
$ n \Delta\,Y\,+\, n(n - 1) \kappa \Delta$.

     \begin{figure}[ht]
    \centering
  \leavevmode
      \includegraphics[width=9cm]{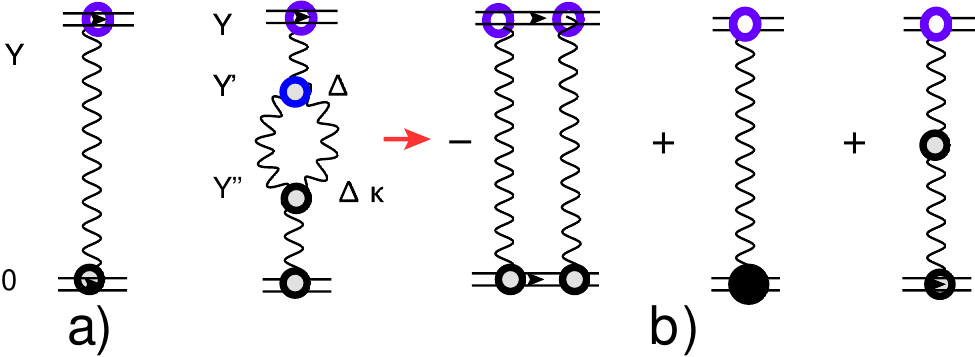}  
      \caption{ The first Pomeron diagrams for the dipole-dipole scattering.\fig{1stdia}-a:  the  Pomeron exchange ($G_\pom(Y)$). \fig{1stdia}-b:The first diagram with the Pomeron interactions which reduces to sum of $G_{2 \pom} $ and $G_{\pom} $ with a different intercept.
    The wavy lines denote the BFKL Pomerons. The open circles show the triple Pomeron vertices. The black large circle denotes $\Lb\Gamma^{2 \pom}_{\pom}\Rb^2 = \kappa^2$. }
\label{1stdia}
   \end{figure}
 


\section{Parton cascade and Pomeron calculus}

In this paper we deal with the parton cascade in which we have two micro processes: the decay of one dipole into two; and the merging of two dipoles  into one.The equation for  probability to have $n$-dipoles ($P_n\Lb Y \Rb$) in this cascade have been studied in Ref.\cite{LELU2} in QCD and they have the form:
\bea \label{PARCSD}
\frac{d P_n(Y)}{d Y} \,\,&=&\,\,\underbrace{\Delta\,\,\,\,\Bigg(-n\,\,P_n(Y)\,\,+\,\,\,(n -1)\,P_{n 
-1}(Y)\Bigg)}_{\mbox{one to two dipoles decay}} \,\nonumber \\
\,&+ & \,\,\Delta\, \kappa\,\underbrace{\Bigg(- \frac{n (n -1)}{2}\,\,P_n(Y) \,\,+\,\,\frac{n (n 
+ 1)}{2}\,\,P_{n+1}(Y)\Bigg)}_{\mbox{two to one dipoles merging}}
\eea
\eq{PARCSD} has a simple structure: for every process of dipole splitting or merging we see two 
terms. The first one with the negative sign  describes a decrease of probability $P_n$ due to the 
process of splitting or merging of dipoles. The second term with  positive sign is  responsible for the 
increase of the probability  due to the same processes of dipole interactions.

The useful tool  for discussing  the structure of the parton cascade is the generating function, which has the following form\cite{MUDI,LELU2}:
\beq \label{Z}
Z\Lb Y, u\Rb\,\,=\,\,\sum_{n}P_n(Y) \,u^n
\eeq
where $P_n(Y) $ is the probability to find $n$ dipoles with rapidity $Y$. For the scattering of one dipole  with the target we have the following initial and boundary conditions:
\beq \label{IC}
\mbox{initial condition:}\,\, Z\Lb Y=0,u\Rb = u; ~~~~~\mbox{boundary  condition:}\,\, Z\Lb Y,u=1\Rb\,=\,1;
\eeq
The boundary condition follows from $P_n(Y)$ being probabilities.

\eq{PARCSD} can be rewritten as the evolution equation for $Z\Lb Y,u\Rb$\cite{LELU2,KOLE} :
:
\beq \label{ZEQ}
\frac{ \partial\,Z\Lb Y,u\Rb}{\partial Y}\,\,=\,\,- \Delta u (1-u) \frac{ \partial\,Z\Lb Y,u\Rb}{\partial u}\,\,+\,\,\h \Delta\,\kappa u (1- u)  \frac{ \partial^2\,Z\Lb Y,u\Rb}{\partial u^2}
\eeq

Introducing $\Y = \Delta\,Y$ we have
 a bit different equation:
 \beq \label{ZEQ1}
\frac{ \partial\,Z\Lb \Y,u\Rb}{\partial \Y}\,\,=\,\,- u (1-u) \frac{ \partial\,Z\Lb \Y, u\Rb}{\partial u}\,\,+\,\,\h \kappa u (1 - u)  \frac{ \partial^2\,Z\Lb \Y, u\Rb}{\partial u^2}
\eeq 
The evolution of  $Z\Lb \Y,u\Rb$ is generated by a non-Hermitian operator, which is
\beq\label{H}
\mathcal{H}\,=\,-\,(1-u)\,\Lb  u  \frac{ \partial}{\partial u}\,\,-\,\,\h \kappa \,u  \frac{ \partial^2}{\partial u^2}\Rb
 \eeq

The BFKL Pomeron calculus is a perturbation theory around the BFKL limit of the Hamiltonian. To eмphаsize this we rewrite the Hamiltonian   of \eq{H} as
  
 \begin{subequations} 
  \bea 
\mathcal{H}\,\,&=&\,t\,\Lb 1 - t\Rb \Lb \frac{\pp}{\pp\, t}\,\,-\,\,\frac{\pp^2}{\pp t^2}\Rb\,\,=\,\,
 \underbrace{ t \frac{\partial}{\partial t}}_{ \mathcal{H}_0} \,-\,  
 \underbrace{ t^2\frac{\partial}{\partial t}\,\,+\,\,t (1-t)\,\,\,\frac{\partial^2}{\partial\,t^2}}_{ \mathcal{H}_I} \label{H1}\\
 \mathcal{H}_I &=& \underbrace{- t^2 \dfrac{\pp}{\pp t} }_{\pom\,\to\,2 \pom;\Gamma^{\pom}_{2 \pom}}\,\,+\,\,  \underbrace{ t \dfrac{\pp^2}{\pp t^2} }_{2 \pom\,\to\,\pom;\Gamma^{2\pom}_{ \pom}} \,\,-\,\, \underbrace{ t^2 \dfrac{\pp^2}{\pp t^2} }_{2 \pom\,\to\,2 \pom;\Gamma^{2\pom}_{2 \pom}} \label{VERT}  \eea 
  \end{subequations}  
 where $t= 1-u$.
Using the variable $t$ rather than $u$ is natural at weak coupling/small rapidities as it  means  the dipole-dipole scattering amplitude which  is small in this regime.  \fig{1stdia} shows the first  Pomeron diagrams for the dipole-dipole scattering. One can see that integrations over $\Y'$ and $\Y''$ reduce the diagram of \fig{1stdia}-b to the zero order contribution to $G_{2 \pom}$ and to $G_\pom$ with a different intercept in comparison with \fig{1stdia}-a. 

It is instructive to calculate the diagramm of \fig{1stdia}-b using  the $\omega$ representation:

\beq \label{PC1}
Z\Lb u=1-t ,\,Y\Rb\,\,=\,\,\int\limits^{\epsilon + i \infty}_{\epsilon - i \infty} \frac{ d \,\omega}{2\,\pi\,i} \,e^{ \omega\,\Y}\,z\Lb t,  \omega\Rb\eeq

In this representation  the contribution of \fig{1stdia}-b has the fornm
\bea \label{PC2}
z\Lb \fig{1stdia}\Rb\,\,& =&\,\,-\frac{1}{\omega - 1} \frac{\kappa}{\omega - 2} \frac{\kappa}{\omega -1} =-\frac{1}{\omega - 1} \kappa\Lb \frac{1}{\omega-2}  - \frac{1}{\omega -1}\Rb \kappa\nn\\
& =& \underbrace{- \frac{\kappa^2}{\omega - 2}}_{G_{2 \pom}} +  \underbrace{ \frac{\kappa^2}{\omega - 1}}_{G_{\pom}}   + \underbrace{ \frac{\kappa^2}{\Lb\omega - 1\Rb^2}}_{G_{\pom} \mbox{with different intercept}}\eea

The first contribution to $G_{\pom}$ leads to the renormalization of the  vertices of interaction with dipoles in \fig{1stdia}-a. The second one is responsible for the new intercept of $G_{\pom}$.

\begin{figure}[ht]
    \centering
  \leavevmode
      \includegraphics[width=9cm]{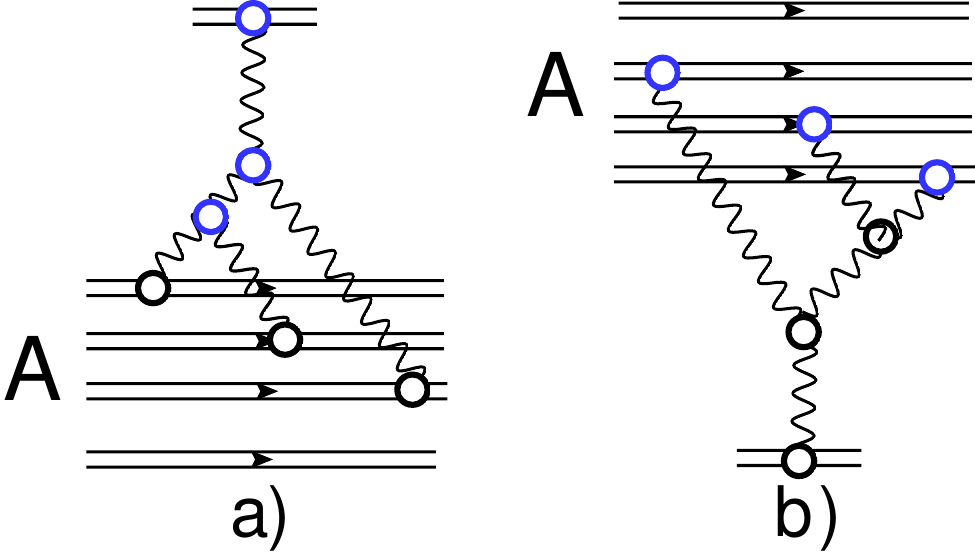}  
      \caption{Fan Pomeron diagrams describing interaction with nucleus. \fig{fandia}-a: a nucleus in the lab r.f.\fig{fandia}-b: a nucleus is fast moving projectile.      
 The wavy lines denote the BFKL Pomerons. The open circles show the triple Pomeron vertices: blue for $\Gamma^{\pom}_{2 \pom}$ and black for $\Gamma^{2 \pom}_{\pom}$. Note, that in the limit $\kappa \to 0$ we need to keep the vertex $\Gamma^{2 \pom}_{\pom} = e^{- \gamma}$ in \fig{fandia}-a  to provide correct BFKL cascade limit.}
\label{fandia}
   \end{figure}

The hamiltonian of \eq{H1} has several attractive features:(i) for $\Gamma^{2 \pom}_{2 \pom}=0$ it coincides with the Braun hamiltonian  for the Pomeron calculus in QCD\cite{BRAUN}; (ii) it gives Balitsky-Kovchegov nonlinear equation in wide range of energy and describe dipole nucleus interaction both in the lab.r.f and in the frame where a nucleus is a fast moving projectile (see \fig{fandia}); and it preserves the probabilistic interpretation of the scattering amplitude given by Mueller and Salam \cite{MUSA}.

 The Pomeron calculus with the hamiltonian of \eq{H1} as well as any other Pomeron calculus satisfies the t-channel unitarity. However we have problem with s-channel unitarity as has been discussed in Ref.\cite{UNRFT} (see section four of this paper). 

\section{Eigenfunctions of the hamiltonian}

\eq{ZEQ1} can be rewritten as the equation for the eigenfunction of 
hamiltonian $\mathcal{H}$ of \eq{H} in the $\omega$ representation of \eq{PC1}:
\beq \label{EF1}
\omega_n z_n\Lb u\Rb \equiv \omega_n z\Psi_n\Lb u\Rb\,\,=\,\,\mathcal{H} \Psi_n\Lb u\Rb
\eeq
Since our hamiltonian is non-hermitian but with real eigenvalues we have two sets of eigenfunctions:  $\Psi_n\Lb u\Rb$  with negative eigenvalues $- \infty \,<\,\tilde{\omega}_n \,<\,0$  and  $\Phi_n\Lb u\Rb$  with  positive  eigenvalues $ +\infty \,<\,\tilde{\omega}_n \,>\,0$. This reflects the two way of describing the scattering amplitude. The first way is the parton model in which the scattering amplitude in the lab.r.f.  is\cite{MUSA}
\beq \label{EF2}
S\Lb Y\Rb\,=\,\sum^{\infty}_{n=1} \,e^{- \kappa\,n} P_n\Lb Y\Rb = \sum B_n(\kappa) \Psi_n\Lb e^{ - \kappa}\Rb e^{\tilde{\omega}_n \,\Y} 
\eeq
In \eq{EF2} we consider the scattering with a dipole and we interpret   $\Delta\,\kappa$ as the scattering two dipoles  amplitude at low energies (se \fig{fandia}). $B_n$ is the coefficient which we need to find from the initial conditions of \eq{IC}.

On the other had we have a different way to describe the scattering amplitude using the factorial moments
\beq \label{MK}
M_k \,=\,\sum^{\infty}_{n=1}\frac{n!}{(n-k)!} P_n\Lb Y\Rb; ~~~\rho_k \equiv \frac{M_k}{k!};
\eeq
In the moments representation the scattering amplitude takes the form (see for example Ref.\cite{LELU2} ):
\beq \label{EF3}
S\Lb Y\Rb\,=\,\sum^{\infty}_{n=k} (- 1)^k\, \rho_k\,\kappa^k\, = \sum C_n(\kappa) \Phi_n\Lb e^{ - \kappa}\Rb e^{\omega_n \,\Y} 
\eeq
Again $C_n$ is the coefficient which we need to find from the initial conditions of \eq{IC}. Series of \eq{EF3} emerges in the Pomeron calculus with  $G_{n \pom} = e^{\omega_n \,\Y} $. This series is basically an asymptotic series which we need to sum. The result of summation should give the analytic function given by \eq{EF2}.

For $\kappa = 0$ our cascade coincides with the BFKL cascade and in Ref.\cite{UTMM} both sets of eigenfunction have been found. It turns out that 
\beq \label{EF4}
\Phi_n^{BFKL}\Lb u\Rb\,=\, \left({u-1\over u}\right)^n;
~~~~~~~~~~~~~~~~~
\Psi_n^{BFKL}\Lb u\Rb\,=\, \left({u-1\over u}\right)^{-n}\eeq
with the corresponding eigenvalues $\omega^{BFKL}_n=n \Delta$ and $\tilde \omega^{BFKL}_n=-n\Delta$.
\eq{EF2} take the following form for the BFKL cascade:
\beq \label{EF5}
Z^{BFKL}\Lb u,  Y \Rb\,=-\sum^\infty_{n=1}\Psi^{BFKL}_n \,e^{-n\,\Delta\, Y}  =1-\frac{1}{1+\frac{u}{1-u}e^{-\Delta Y}}=\frac{1}{1-(1-\frac{1}{u}) e^{\Delta Y}}:~~ S\Lb \Y\Rb= Z^{BFKL}\Lb u=e^{- \gamma},  Y \Rb;
\eeq

 $\gamma$ is the interaction of two dipoles at low energies. Note, that for $\kappa=0$ we have to introduce this amplitude (see for example \fig{fandia}-a) . We need to keep $\Delta \kappa = \gamma$ at $\kappa \to 0$ to have  dipole nucleus scattering in this case.

For \eq{EF3} we have:
\beq \label{EF6}
Z^{BFKL}\Lb u,  Y \Rb\,=-\sum^\infty_{n=1}\Phi^{BFKL}_n \,e^{n\,\Delta\,Y}  =\frac{1}{1-(1-\frac{1}{u}) e^{\Delta Y}}:~~~~~ S\Lb \Y\Rb= Z^{BFKL}\Lb u=e^{- \gamma},  Y \Rb;
\eeq
We got the same answers for two series. However, series of \eq{EF6} is asymptotic one at large $Y$  which has been summed using Borel resummation procedure, while \eq{EF5} is an absolutely converged series at large $Y$. It is instructive to note  at small $Y$ gives the contribution of the exchange of one Pomeron while the series of \eq{EF5} has to be resummed.

\subsection{Eigenfunction for negative eigenvalues}


\eq{EF1} can be rewritten in the Sturm-Liouville  form\cite{POLY}:
\beq \label{EFNV1}
\frac{d}{d\,u}\Lb p(u) \frac{d z\Lb \omega, u\Rb}{ d\,u} \Rb\,= - \lambda s(u) z\Lb \omega, u\Rb 
\eeq
with
\beq \label{EFNV2}
p\Lb u \Rb\,=\,\h \kappa e^{ - \frac{2 \,u}{\kappa}} ;~~~s\Lb u\Rb\,=\,\frac{2}{\kappa} \frac{1}{u ( 1 - u)} e^{ - \frac{2 \,u}{\kappa}};~~~~~\omega\, =\, - \lambda;
\eeq

  The Sturm- Liouville equation of \eq{EFNV1} has the following general features\cite{KOLEV,POLY}:
  \begin{enumerate}
  \item\quad It has infinite set of eigenvalues $\lambda_n$, which monotonically increase with $n$ . In our case of \eq{EFNV1} all $\lambda_n > 0$. 
   \item\quad The multiplicity of each eigenvalue is equal to 1.
   \item \quad The eigenfunctions $\Psi_n(u)$ are orthogonal 
   \beq \label{GS2}
   \int^1_0\!\!
  \!du\,s(u) \,\Psi_n(u)\,\Psi_m(u)\,\,=\,0 ~~~\mbox{for}~~n \neq n
   \eeq
  \item \quad For large $n$ 
  \beq \label{GS3}
  \lambda_n\,\,=\,\,\frac{\pi^2\,n^2}{\delta^2}\,+\,\mathcal{O}\Lb 1 \Rb ~~\mbox{with}~~\delta = \int^1_0 \!\!\!d u \sqrt{\frac{s(u)}{p(u)}}
  \eeq
     \item \quad  For our equation
     \beq \label{GS4}
     \delta =\sqrt{\frac{2}{\kappa}} \int^1_0 \frac{ d u}{\sqrt{u(1-u)}} \,=\,\,\,\pi  
\sqrt{\frac{2}{\kappa}}\eeq
    leading to
     \beq \label{GS5}
     \lambda_n = \,\,\h\kappa\,n^2
     \eeq
     
     \item \quad An arbitrary function F(u), that has a continuous derivative and satisfies the boundary conditions of the Sturm-Liouville problem (in other words, a function that we, as physicists, want to find), can be expanded into absolute and uniformly convergent series in eigenfunctions:   
    \beq \label{GS6}
    F\Lb u\Rb \,=\,\sum^{\infty}_{n=1} F_n \Psi_n\Lb u\Rb ~~\mbox{with}~~F_n = \frac{1}{||\Psi_n||}\intl^1_0 d u' s\Lb u'\Rb F\Lb u'\Rb \Psi_n\Lb u'\Rb~~\mbox{where}~~~||\Psi_n|| \,=\, \intl^1_0 d u' s(u')\Psi_n^2\Lb u'\Rb
    \eeq  
    
    \item \quad For $n \gg 1$  $\Psi_n(u)$ takes the form:
     \beq \label{GS7} 
    \frac{ \Psi_n\Lb u\Rb}{||\Psi_n||} =\Lb \frac{4}{\delta^2\, p(u)\, s(u)}\Rb^{\frac{1}{4}}\sin\Lb \frac{\pi\,n}{\delta}\intl^u_0 d u'\sqrt{\frac{s(u')}{p(u')}} \Rb\,\,+\,\,\mathcal{O}\Lb\frac{1}{n}\Rb
   \eeq  
     \end{enumerate}
     Therefore, one can see that  we know a lot about $\Psi_n$ which corresponds to the negative eigenvalues. In Ref.\cite{KOLEV} $\Psi_n$ are found for the arbitrary values of $n$.  This solution is 
     \beq \label{EFNV3}     
     \Psi_n\Lb u\Rb\,\,=\,\,e^{\frac{1 - v}{2\,\kappa}}\sqrt{1 - v^2} \,S_{n,1}\Lb v\Rb
     \eeq
     with $v = 1 - 2\,u$. 
     $S_{n,1}\Lb v\Rb   $  is
  the prolate spheroidal wave function with the fixed
order parameter $m \,\,=\,\,1$ and arbitrary degree parameter $n$ which is an
integer, $S_{n,m}(c,v)$  (see
Refs. \cite{ABST,SPHF}),
namely,
\beq \label{EFNV4}
\frac{d}{d v}\,\left((1 - v^2)\,\frac{d S_{n,m}(c,v) }{d
v}\right)\,\,+\,\,\left(\,\lambda^m_n\,\,-\,\, c^2 \,v^2\,\,-\,\,\frac{m^2}{1 - v^2}
\right)\,\,\,S_{n,m}(c,v)\,\,=\,\,0
\eeq
   \eq{EFNV1}   gives the following parameters of \eq{EFNV4}:
   \beq \label{EFNV5}
   \lambda_{n,1} = \frac{1}{4 \kappa^2} \,-\,\frac{2 \tilde\omega_n}{\kappa} ;~~
   \tilde \omega_n = -\h \kappa \lambda_{n,1} + \frac{1}{ 8\,\kappa};~~~m = 1;~~ c = \frac{1}{2 \kappa};
   \eeq
 For  large $n$ $ \lambda_{n,1}    \,\,\to\,\,n (n+1)$ while for small $n$ $ \lambda_{n,1}    \,\,\to\,\,n/\kappa $ (see Ref.\cite{ABST} formulae {\bf 21.6.2} and {\bf 21.7.6}).

 ~

  \begin{boldmath}
\subsection{$P_n$ for $\kappa \ll 1$}
\end{boldmath}

 In this subsection we are going to find $P_n$ of \eq{PC1} for small $\kappa$. We hope to obtain a simpler set of the eigenvalues and eigenfunctions that in a general case that has been considered in the previous section. Second, we expect a transparent relation between these eigenvalues and the positive eigenvalues that we will consider in the next subsection.
 
 In the omega representation of \eq{PC1} the set of equations of \eq{PARCSD} takes the following form, taking into account \eq{IC}:
  \begin{subequations} 
  \bea     
    && \Lb \omega + 1\Rb P_1\Lb \omega\Rb\,\,-\,\,\kappa \,P_2\Lb \omega\Rb   \,=\,1;\label{PN1}\\
    &&  \Lb \omega +2\Rb P_2\Lb \omega\Rb\ \,=\,P_1\Lb \omega\Rb \,+ \kappa \Lb - P_2\Lb \omega\Rb \,+\,3 \,P_2\Lb \omega\Rb\Rb;\label{PN2}\\
    &&  \Lb \omega +3\Rb P_3\Lb \omega\Rb\ \,=\,2\,P_2\Lb \omega\Rb \,+ \kappa \Lb - 3P_3\Lb \omega\Rb \,+\,6 \,P_4\Lb \omega\Rb\Rb;\label{PN3}\\ 
  &&  \Lb \omega +4\Rb P_3\Lb \omega\Rb\ \,=\,3\,P_3\Lb \omega\Rb \,+ \kappa \Lb - 6\,P_4\Lb \omega\Rb \,+\,10 \,P_5\Lb \omega\Rb\Rb;\label{PN4}\\ 
&&\Lb \omega + n \Rb \,P_n\Lb \omega\Rb \,\,= \,\,(n -1)\,P_{n 
-1}\Lb \omega\Rb\,+ \,\,\kappa\,\Lb- \frac{n (n -1)}{2}\,\,P_n\Lb \omega\Rb \,\,+\,\,\frac{n (n 
+ 1)}{2}\,\,P_{n+1}\Lb \omega\Rb\Rb ;\label{PNN}  
       \eea
    \end{subequations} 
    Note that the r.h.s., of \eq{PN1} comes from the initial condition of \eq{IC}. 
    
    Plugging \eq{PN2} into \eq{PN1} we obtain:
    \beq \label{PN5}
    \Lb \omega + 1\Rb P_1\Lb \omega\Rb\,\,-\,\,\kappa \frac{P_1\Lb \omega\Rb \,\,+\,\,\mathcal{O}\Lb \kappa\Rb}{\omega\,+\,2}  \,=\,1 ~
 \eeq
 Solution to \eq{PN5} has the form  for small values of $\kappa$:
 \beq \label{PN6}
 P_1\Lb \omega\Rb \,\,=\,\,\frac{1 + \mathcal{O}\Lb \kappa\Rb}{\omega\,\,+\,\,1\,\,-\,\,\kappa}
 \eeq 
 For $P_2   -  P_1$ from \eq{PN1} and \eq{PN2} we have
 \beq \label{PN7}
( \omega  + 2)\Lb P_2 - P_1\Rb\,\,=\,\,-1\,\,+\,\,\kappa \Lb - 2 P_2 + 2 P_3\Rb
 \eeq
 
 At $\kappa = 0$ we see that
 \beq \label{PN8}
 \Lb P_2 - P_1\Rb\,\,=\,\,- \frac{1}{\omega\,+\,2} 
 \eeq  
 which gives $\Lb P_2\Lb \Y\Rb  - P_1\Lb \Y\Rb\Rb = -  e^{ -\,2 \Y} $. In other word both $P_1\Lb \Y\Rb  =  e^{ -\, \Y}  $ and  $C_2\Lb \Y\Rb\,=\,\Lb P_2\Lb \Y\Rb  - P_1\Lb \Y\Rb\Rb $ diagonalize the matrix of $P_n$. The general expression for $C_n$ is
  \beq \label{PN9} 
 C_n \,\,=\,\,\sum^{n-1}_{k=0} (-1)^k\,\frac{(n-1)!}{k!\,(n-1 -k)!}\,P_{k+1} 
 \eeq
 \eq{PN7} leads to
 \beq \label{PN10}
C_2\Lb \omega\Rb\,\,=\,\, \Lb P_2 - P_1\Rb\,\,=\,\,\frac{ -1\,\,+\,\,\mathcal{O}\Lb \kappa\Rb}{\omega\,\,+\,\,2\,\,-\,4\,\kappa}
\eeq
if we plug in this equation $P_3 = \frac{2\,P_2}{\omega + 3}$ from \eq{PN3}.
 
 From the general \eq{PNN} we can get the following equation\footnote{ In the next section we will give the more expanded discussions how to get such an equation.} 
for 
 \beq \label{PN11}
 \tilde{\omega}_n\,\,=\,\,-n\,\,+\,\,\kappa\,n^2 
 \eeq

 The first glance at this equation shows a striking difference with \eq{EFNV5}.
 However, for small values of $\kappa$ $\lambda_{n,1}$ leads to \eq{PN11}  (see Ref.\cite{ABST} formulae {\bf 21.7.6}).    
 
  ~

\subsection{Positive eigenvalues}

 
 We are going to find the positive eigenvalues by writing the equations for the factorial moments of \eq{MK} using that (e.g. see Ref.\cite{LELU2} )
 \beq \label{PEV1} 
 \rho_n\,\,=\,\,\frac{1}{n!} \frac{ \pp^n}{\pp \,u^n} Z\Lb \Y,u\Rb\Big{|}_{u=1}
 \eeq
 Differentiating \eq{ZEQ1} we obtain
  \begin{subequations} 
  \bea  
  \frac{\pp \rho_1\Lb \Y\Rb}{\pp\,\Y}\,\,&=&\,\,\rho_1\Lb\Y\Rb \,\,-\,\,\kappa\,\rho_2\Lb y\Rb; \label{PEVM1} \\ 
   \frac{\pp \rho_2\Lb \Y\Rb}{\pp\,\Y}\,\,&=&\,\,\,\rho_1\Lb\Y\Rb \,\,+\,\,2\Lb 1 - \h \kappa\Rb\rho_2 \Lb \Y\Rb\,\,-\,\,3 \kappa\,\rho_3\Lb \Y\Rb; \label{PEVM2} \\ 
   \frac{\pp \rho_3\Lb \Y\Rb}{\pp\,\Y}\,\,&=&\,\,2\,\rho_2\Lb\Y\Rb \,\,+\,\,3\Lb 1 -  \kappa\Rb\rho_3 \Lb \Y\Rb\,\,\,\,\,\,\,-\,\,6 \kappa\,\rho_4\Lb \Y\Rb; \label{PEVM3}  \\  
    \frac{\pp \rho_4\Lb \Y\Rb}{\pp\,\Y}\,\,&=&\,\,3\,\rho_3\Lb\Y\Rb \,\,+\,\,4\Lb 1 - \frac{3}{2} \kappa\Rb\rho_4 \Lb \Y\Rb\,\,\,-\,\,10 \kappa\,\rho_5\Lb \Y\Rb; \label{PEVM4}  \\
  \frac{\pp \rho_{n-1}\Lb \Y\Rb}{\pp\,\Y}\,\,&=&\,\,(n-2)\,\rho_{n-2}\Lb\Y\Rb \,\,+\,\,(n - 1)\Lb 1 -  \h(n-2) \kappa\Rb\rho_{n-1}\Lb \Y\Rb\,\,\,-\,\,\h n (n-1)  \kappa\,\rho_{n}\Lb \Y\Rb; \label{PEVM1N} \\  
\frac{\pp \rho_n\Lb \Y\Rb}{\pp\,\Y}\,\,&=&\,\,(n-1)\,\rho_{n-1}\Lb\Y\Rb \,\,+\,\,n\Lb 1 -  \h(n-1) \kappa\Rb\rho_n\Lb \Y\Rb\,\,\,-\,\,\h n (n+1)  \kappa\,\rho_{n+1}\Lb \Y\Rb; \label{PEVMN}  \\
  \frac{\pp \rho_{n+1}\Lb \Y\Rb}{\pp\,\Y}\,\,&=&\,\,n\,\rho_{n}\Lb\Y\Rb \,\,+\,\,(n+1)\Lb 1 -  \h n\kappa\Rb\rho_{n+1}\Lb \Y\Rb\,\,\,-\,\,\h (n+2) (n+1)  \kappa\,\rho_{n+2}\Lb \Y\Rb; \label{PEVMN1}   \eea
     \end{subequations}

     For $\kappa \to 0$ we have the equations for the moments in the BFKL parton cascade. Solutions for this case are known(e.g. Ref.\cite{KHLE}):
   \beq \label{PEV2}
   \rho^{BFKL}_n\Lb \Y\Rb\,\,=\,\,e^{\Y}\Lb e^{\Y} \,-\,1\Rb^{n-1};~~~~~  
      \rho^{BFKL}_n\Lb \omega\Rb  \,=\,(n-1)!\prod^n_{k=1}\frac{1}{\omega - k};
 \eeq
 The $\omega$ representation is defined in \eq{PC1}.    Recall that the exchange of the BFKL Pomeron is $G_{\pom} = e^{\Y}$ with our definition of $\Y$. 
          
    The set of power moments $c_k =\frac{1}{k!} \sum^\infty_{n=1} n^k \,P_n\Lb\Y \Rb$      diagonalizes  \eq{PEVM1} - \eq{PEVMN1}. Each of them is equal to $c_k\,=\,\frac{1}{\omega\,-\,k} $ or $e^{k \,\Y}$ in $\Y$ representation. On the other hand
    \beq \label{PEV3} 
    c_n\Lb \omega\Rb\,=\,\sum^{n-1}_{k=0} \frac{(n-1)!}{k!\,(n-1 -k)!}\,\rho_{k+1};~~~~~~
     \rho_n\Lb \omega\Rb\,=\,\sum^{n-1}_{k=0}(-1)^k \frac{(n-1)!}{k!\,(n-1-k)!}\,c_{k+1} ;   
    \eeq
    
    Now we attempt to solve  \eq{PEVM1} - \eq{PEVMN1}, considering $\kappa \ll 1$. First, we rewrite these equations in $\omega$ representation (see \eq{PC1}). \eq{PEVM1}  and   \eq{PEVM2} takes the form:
     \begin{subequations} 
  \bea     
    && \Lb \omega - 1\Rb \rho_1\Lb \omega\Rb\,\,+\,\,\kappa \,\rho_2\Lb \omega\Rb   \,=\,1;\label{RHO1}\\
    &&  \Lb \omega - 2\Lb 1 - \h \kappa\Rb\Rb \rho_2\Lb \omega\Rb\ \,=\,\rho_1\Lb \omega\Rb \,-\,3 \kappa \rho_3\Lb \omega\Rb;\label{RHO2}
    \eea
    \end{subequations} 
    Note that the r.h.s., of \eq{RHO1} comes from the initial condition of \eq{IC}. Plugging $\rho_2$ from the second equation we obtain:
    \beq \label{RHO3}
    \Lb \omega - 1\ \,\,+\,\,\kappa \frac{1}{ \omega - 2\Lb 1 - \h \kappa\Rb} \Rb \rho_1\Lb \omega\Rb   \,=\,1    -3 \kappa^2\frac{1}{ \omega - 2\Lb 1 - \h \kappa\Rb} \rho_3\Lb \omega\Rb
    \eeq   
    
    Neglecting corrections of $\kappa^2$ size, one can see that solving \eq{RHO3} we obtain:
     \beq \label{RHO4} 
   \rho_1\Lb \omega\Rb \,\,=\,\,c_1\Lb \omega\Rb\,\,=\,\,\frac{1}{\omega \,-\,1\,-\,\kappa}
   \eeq
    
    It is easy to see that \eq{RHO4} corresponds to the third contribution  of \fig{1stdia}-b in the Pomeron calculus. Summing \eq{RHO1} and \eq{RHO2} we obtain the following equation for $c_2 = \rho_2+\rho_1$:
    \beq \label{RHO5}
    \Lb \omega - 2 + 2 \kappa\Rb c_2=  1  \,+\, 2 \kappa \rho_1\, -\, 3 \kappa \rho_3
 \eeq
 Plugging in this equation $\rho_3$ from \eq{PEVM3} ;
 \beq \label{RHO6}
    \rho_3 = \frac{ 2 \Lb c_2 - c_1\Rb}{\omega - 3 (1 - \kappa) } \,\,+\,\,\mathcal{O}\Lb \kappa^2\Rb
    \eeq
    we have
     \beq \label{RHO6}   
     \Lb \omega\,-\,2\,+ 2 \kappa + \frac{ 6 \kappa}{ \omega - 3 (1 - \kappa) } \Rb c_2= 1 \,+\,2 \kappa c_1 + \frac{ 6\kappa}{ \omega - 3 (1 - \kappa) }  c_1
     \eeq  
     
     In the vicinity of $\omega = 2\, +4\,\,\kappa$  \eq{RHO6} takes the form:
      \beq \label{RHO6}  
      c_2\Lb \omega \Rb\,\,=\,\, \frac{1\,\,-\,\,4 \kappa}{ \omega\,\,-\,\,2\,\,-\,\,4\,\kappa} \,\,+\,\,\mathcal{O}\Lb \kappa^2\Rb
      \eeq
      
\begin{figure}[ht]
    \centering
  \leavevmode
      \includegraphics[width=15cm]{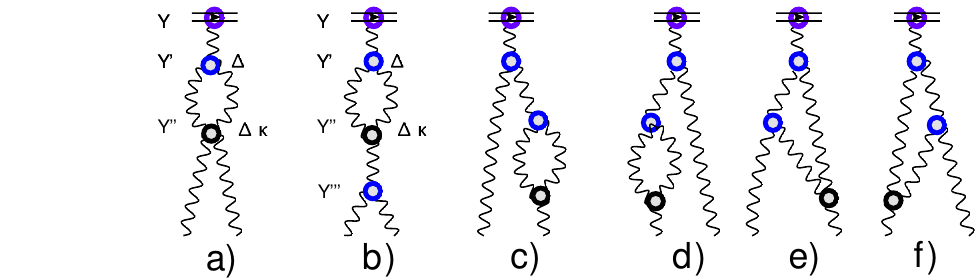}  
      \caption{ The  Pomeron diagrams for $\rho_2$ in the dipole-dipole scatteringof the order of  $\kappa$.
    The wavy lines denote the BFKL Pomerons. The open circles show the triple Pomeron vertices.  }
\label{rho2dia}
   \end{figure}

    The calculations of  \eq{PC2} -type shows that sum of the Pomeron diagrams of  \fig{rho2dia} leads to     \eq{RHO6}.

      We need to sum \eq{PEVM1}-\eq{PEVM3} for finding $c_3=\rho_3 + 2 \rho_2 + \rho_1$ and express $\rho_4$ from \eq{PEVM4} neglecting terms proportional to $\kappa^2$. The resulting equation takes the form:
      \beq \label{RHO7}  
     \Lb \omega\,-\,3\,\,-\,\, 9 \kappa\Rb c_3(\omega) = 1 \,-\,21 \kappa \, c_2(\omega) \,+\,12\,\kappa\,c_1\Lb \omega\Rb
     \eeq
     which leads to
     \beq \label{RHO8}
     c_3\Lb \omega\Rb\,\,=\,\,\frac{1\,\,-\,21 \kappa  c_2(\omega) \,+\,12\,\kappa\,c_1\Lb \omega\Rb}{\omega\,\,-\,\,3\,\,-\,\,9 \,\kappa} \,\,=\,\,     
   \frac{1\,\,-\,21 \kappa  c_2(\omega=3) \,+\,12\,\kappa\,c_1\Lb \omega=3\Rb}{\omega\,\,-\,\,3\,\,-\,\,9 \,\kappa}  \,\,=\,\,\frac{1\,\,-\,\,15\,\kappa }{\omega\,\,-\,\,3\,\,-\,\,9 \,\kappa}   
  \eeq
         
We can easily calculate the intercept for $c_n$ using \eq{PEVM1N}-  \eq{PEVMN1}.    From \eq{PEV3} one can see that $c_n = \rho)n + (n-1)\,c_{n-1} + \dots$ . Multiplying \eq{PEVM1N} by $(n - 1)$ and adding to \eq{PEVMN} we obtain:
     \beq \label{RHO9}
     \Lb \omega - n -\h \kappa n (n-1)^2 - \h n (n-1) \,-\, \frac{\h \kappa n^2 (n+1)}{\omega \,-\,n - \mathcal{O}\Lb \kappa\Rb} \Rb c_n = 1 \,+\,\mathcal{O}\Lb \kappa\Rb
\eeq
which leads to

     \bea \label{RHO10}
  c_n\,\,&=&\,\,\frac{1 + \mathcal{O}\Lb \kappa\Rb}{ \omega\,-\,n \,+\,\kappa\Lb \h n(n-1)^2 + \h n (n-1) \,+\,\frac{ \h n^2(n+1)}{\omega - n- 1  - \mathcal{O}\Lb \kappa\Rb} \Rb}  \,\,\nn\\
  &=&\,\,\frac{1 + \mathcal{O}\Lb \kappa\Rb}{ \omega\,-\,n \,+\,\kappa\Lb \h n^2(n-1)^2  \,-\,\h n^2(n+1) \Rb}\,=\,\frac{1 + \mathcal{O}\Lb \kappa\Rb}{ \omega\,-\,n \,-\,\kappa\,n^2 }  
  \eea
where  we put $\omega = n $ in the second equation. From this equation one can see that we obtain the positive eigenvalues  which are equal to
  \beq \label{RHO11}      
    \omega_n\,\,=\,\,n\,\,+\,\,\kappa\,n^2
    \eeq 
   
   For finding 
    the contribution $\mathcal{O}\Lb \kappa\Rb$ in \eq{RHO9}  we use $c_n$ from \eq{PEV3}. The equation for $C_n$ takes the form:
     \bea \label{RHO12}                
 &&     (\omega\,-n)c_n\,\,=\,\,1\,\,\,-\h \kappa\sum^{n-1}_{k=0} C^k_{n-1} \,k\,(k+1) \rho_{k+1} \,\,-\,\, \h \kappa\sum^{n-1}_{k=0} C^k_{n-1} \,(k+1)\,(k+2) \rho_{k+2}\nn\\
&&=\,\,1\,\,\,-\h \kappa n(n-1) \rho_n - \h\kappa \sum^{n-2}_{k=0} C^k_{n-1} \,k\,(k+1) \rho_{k+1}   \,-\,\h \kappa n(n+1)\rho_{n+1} \nn\\
&&- \h \kappa(n-1)^2 n\rho_n 
    -\h \kappa\sum^{n-3}_{k=0} C^k_{n-1} \,(k+1)\,(k+2)\rho_{k+2}       
    \eea    
    where $C^k_{n-1}$ are binomial coefficients.

  Plugging in this equation $\rho_{n+1} $ from \eq{PEVMN1} we obtain
     \bea \label{RHO13}                
     (\omega\,-\,n\,- \kappa n^2)  c_n\,\,
&=&\,\,1\,\,\,- \h\kappa \sum^{n-2}_{k=0} C^k_{n-1} \,k\,(k+1) \rho_{k+1}(\omega = n)\nn\\  
   & -&\h \kappa\sum^{n-3}_{k=0} C^k_{n-1} \,(k+1)\,(k+2)\rho_{k+2}(\omega = n  )\,-\,\kappa\,n^2   \sum^{n-2}_{k=0} C^k_{n-1} \rho_{k+1}(\omega = n )
    \eea   
  Neglecting terms that are proportional to $\kappa^2$ we can use \eq{PEV2} for $\rho_{k+1} (\omega = n)$. Summing over $k$ we obtain:
   \beq \label{RHO14}                
     (\omega\,-\,n\,- \kappa n^2)  c_n\,\,
=\,\,1\,\,-\,\h  \kappa \Lb n +2\Rb \,n\, \Lb n - 1\Rb   \eeq

  Note, that \eq{RHO14} describe $c_1$,$c_2$ and $c_3$ calculated above.

    Comparing \eq{RHO11} with      \eq{PN11} one can see that
     \beq \label{RHO15}        
     \tilde{\omega}_n = \omega_{-n} 
     \eeq

 ~

\section{Scattering amplitude}

\subsection{Sum of decreasing exponents}
 First we consider the representation of $S\Lb \Y\Rb$ given by \eq{EF2}, viz.:
 \beq \label{SA1}
 S\Lb Y\Rb\, =\,Z\Lb \Y, u=e^{ - \kappa}\Rb\,=   \sum B_n(\kappa) \Psi_n\Lb e^{ - \kappa}\Rb e^{\tilde{\omega}_n \,\Y}  
 \eeq
 where $\Psi_n$ are given by \eq{EFNV3}.   We need first to find the asymptotic solution to \eq{ZEQ1} which does not depend on $\Y$ and has the following boundary conditions:
 
 \beq  \label{SABC}
 Z^{asymp} \Lb u=0\Rb \,\,=\,\,0;~~~~ Z^{asymp} \Lb u=1\Rb \,\,=\,\,1;
\eeq
 
 One can see that
  \beq  \label{SA2}
\displaystyle{ Z^{asymp} \Lb u \Rb \,\,=\,\,\frac{  1\,\,-\,\,e^{ - \frac{2}{\kappa}\,u}}{  1\,\,-\,\,e^{ - \frac{2}{\kappa}}} }
\eeq

$Z\Lb \Y,u\Rb - Z^{asymp}\Lb u\Rb$  has the following initial conditions:

\beq  \label{SAIBC}
Z\Lb \Y=0,u\Rb -  Z^{asymp} \Lb u\Rb \,\,=\,\,u - Z^{asymp} \Lb u=\Rb \eeq

Therefore, 
\beq \label{SA3}
Z\Lb \Y,u\Rb -  Z^{asymp} \Lb u\Rb\,\,=\,\,\sum^{\infty}_{n=1} B_n e^{\frac{1 - v}{2\,\kappa}}\sqrt{1 - v^2} \,S_{n,1}\Lb \frac{1}{2 \,\kappa}, v\Rb  e^{ - \h \,\kappa \lambda_{n,1}\,\Y}
\eeq
where 
\beq \label{SA4}
B_n\,\,=\,\,\frac{1}{||S(n,1)||}\intl^1_{-1}d v' \frac{\Lb \h(1 - v') -Z^{asymp} \Lb v'\Rb\Rb}{\sqrt{1 - v'^2}}e^{\frac{-1 + v'}{2\,\kappa}}S_{n,1}\Lb \frac{1}{2 \,\kappa}, v'\Rb~~\mbox{with}~~||S(n,1)||=\frac{2 n (n + 1)}{2 n  +  1};
\eeq

~

\subsection{Pomeron calculus}
From \eq{PEV1} one can see that
\beq \label{POC1}
Z\Lb\Y, u\Rb = \sum_n \rho_n\Lb \Y\Rb (1 - u)^n;~~~~~S\Lb \Y\Rb = Z\Lb \Y, u=e^{ - \kappa}\Rb;
\eeq
However, it is more convenient for us to use the expression for the scattering amplitude which follows directly from \eq{EF2}:
\beq \label{POC2}
S\Lb \Y\Rb \,=\,\sum^{\infty}_{n=1} \,e^{- \kappa\,n} P_n\Lb Y\Rb=
\,\sum^{\infty}_{n=1}P_n\Lb Y\Rb\sum^\infty_{k=0} \frac{(-\kappa)^k}{k!}\,n^k=
\sum^\infty_{k=0} \frac{(-\kappa)^k}{k!}\sum^\infty_{n=1} n^k\,P_n\Lb Y\Rb=
\sum^\infty_{k=0} (- \kappa)^k c_k\Lb \Y\Rb
\eeq
Plugging \eq{RHO14} into this equation we have
\beq \label{POC3}
S\Lb \Y\Rb \,=\sum^\infty_{k=0} (- \kappa)^k \Lb 1\,- \h \kappa (k+2) k (k-1)\Rb
e^{ \Lb k +   \kappa\,k^2\Rb\,\Y}
\eeq
\eq{POC3} is the series of the Green's function of the many Pomerons exchanges each of which is equal to $G_{k \pom} = e^{ \Lb k +   \kappa\,k^2\Rb\,\Y}$.

This series is not only asymptotic one but it cannot be summed using Borel approach. Actually in Ref.\cite{UTMM} it is proposed the way how to sum such series (see appendix A of this paper). The prescription is to replace
$e^{ \Lb k +   \kappa\,k^2\Rb\,\Y}$ by
\beq \label{POC4}
e^{ \Lb k +   \kappa\,k^2\Rb\,\Y}\,\,=\,\,e^{ (1 - \kappa) \,k\,Y}\intl^\infty_{-\infty}
\frac{d \lambda}{ 2 \sqrt{\pi\,\kappa \,\Y} } \exp\Lb - \frac{ \Lb \kappa  \Y - \lambda\Rb^2}{ 4 \kappa  \Y} \,\,-\,\,\lambda\,k\Rb
\eeq

Plugging into \eq{POC3} this representation we reduce the series to the Borel summed one. Indeed,  \eq{POC3} takes the form:
\beq \label{POC5}
S\Lb \Y\Rb \,=\intl^\infty_{-\infty}
\frac{d \lambda}{ 2 \sqrt{\pi\,\kappa \,\Y} } \exp\Lb - \frac{ \Lb \kappa  \Y - \lambda\Rb^2}{ 4 \kappa  \Y} \Rb
\sum^\infty_{k=0} (- \kappa)^k \Lb 1\,- \h \kappa (k+2) k (k-1)\Rb
e^{ \Lb 1 - \kappa  - \lambda   \Rb\,k\,\Y}
\eeq
Summing over $k$ we obtain
\beq \label{POC6}
S\Lb \Y\Rb \,=\intl^\infty_{-\infty}
\frac{d \lambda}{ 2 \sqrt{\pi\,\kappa \,\Y} } \exp\Lb - \frac{ \Lb \kappa  \Y - \lambda\Rb^2}{ 4 \kappa  \Y} \Rb\Bigg\{-\frac{\zeta ^2 \kappa ^3}{(\zeta  \kappa +1)^3}-\frac{3 \zeta ^2 \kappa ^3}{(\zeta  \kappa +1)^4}+\frac{1}{\zeta  \kappa +1}\Bigg\}
\eeq
with $\zeta = e^{ \Lb 1 - \kappa \Rb\,\,\Y\,\,-\,\,\lambda}$.

\begin{figure}[ht]
    \centering
  \leavevmode
      \includegraphics[width=12cm]{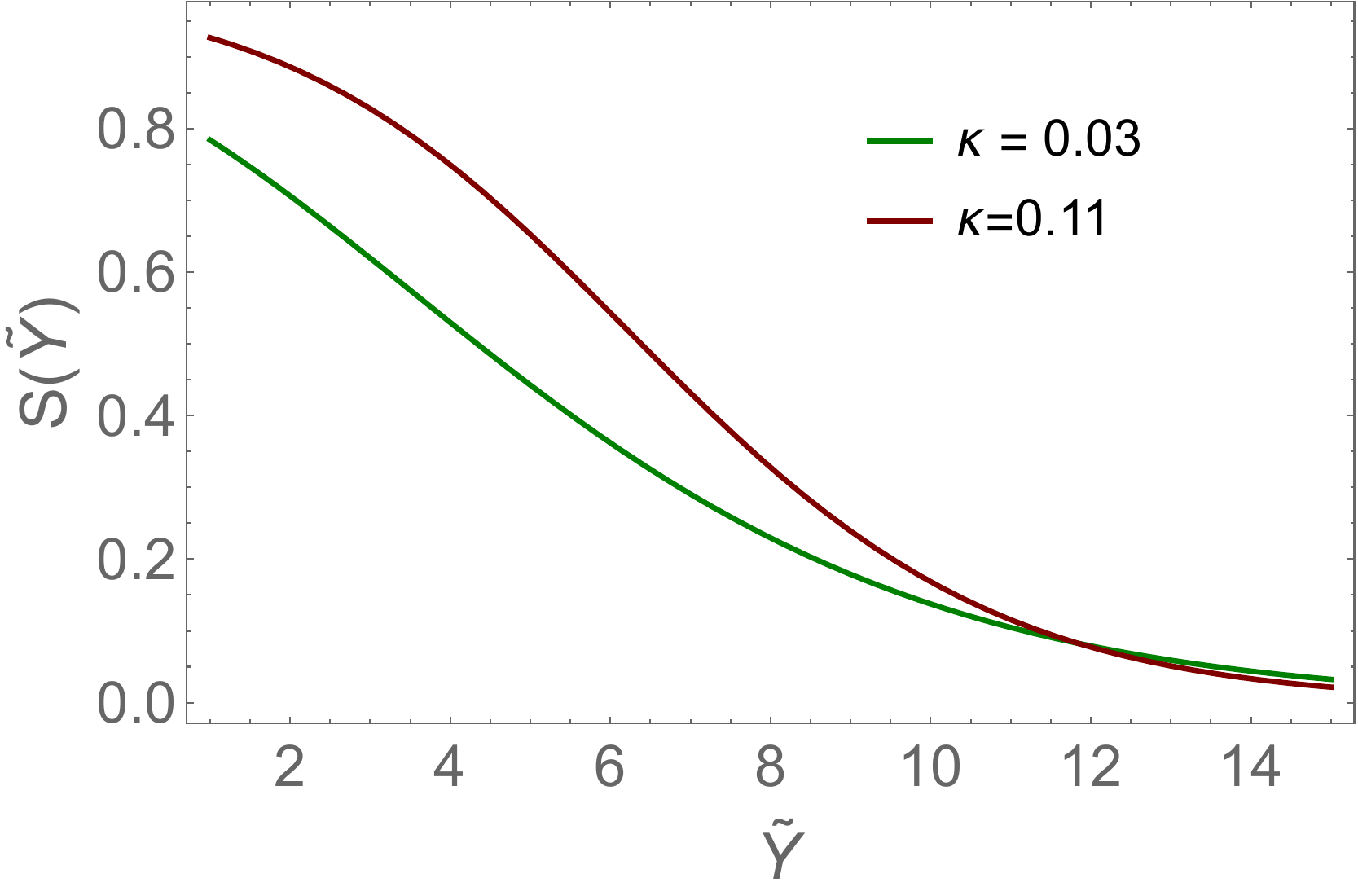}  
      \caption{ The scattering matrix $S\Lb \Y\Rb$ of \eq{POC6} versus $\Y$ at different values of $\kappa$. }
\label{ss}
   \end{figure}

In \fig{ss} we plot $S\Lb \Y\Rb$ versus $\Y$ at two different values of $\kappa$, which we will discuss below. One can see that $S\Lb \Y\Rb$ decreases at large $\Y$ certifying  that our way of resummation of the asymptotic series gives the analytic function that describes the scattering amplitude.

At large $\Y$ \eq{POC6} leads to
\beq \label{POC7}
S\Lb \Y\Rb \xrightarrow{\Y \gg 1} \Lb \frac{1}{\kappa} - 1\Rb e^{- (1 - 3 \kappa) \Y} 
\eeq
     
~


\subsection{The value of $\mathbf{\kappa}$.}
We have learnt three lessons. First, the Pomeron calculus with the hamiltonian of \eq{H1} leads to the intercepts for n Pomeron exchanges which are $\omega_n = n + \kappa n^2$. Second, we found the way to sum the asymptotic series for the scattering amplitude and, third, this amplitude shows the saturation at high energies ($ S \Lb \Y\Rb \xrightarrow{\Y\gg 1} \,0$).

As we have mentioned the hamiltonian of \eq{H1}  without vertex $ 2 \pom \to 2 \pom$ (see \eq{VERT})  is the one dimensional version of the Braun hamiltonian in QCD \cite{BRAUN}. This hamiltonian is the minimum that we need to get the BK equation to be valid  for the deep inelastic scattering in the Bjorken frame in which the virtual photon is at the rest. Note that the first nonlinear equation which coincides with the BK one in the momentum representation, was derived in this frame\cite{GLR}.  In the framework of this approach vertex $ 2 \pom \to 2 \pom$ from  the additional requirement that the hamiltonian will describe the parton cascade (see Ref.\cite{LELU2} ).  In such interpretation $\Delta \kappa$ corresponds to the QCD $2 \pom \to \pom$ vertex which is proportional to $\bas^2/N^2_c$. Hence $\kappa \propto \Delta/N_c^2$.  In \fig{ss} we take $\kappa = 0.25/9 = 0.03$.

However, the most interesting case for us in  this paper is to interpret the hamiltonian of \eq{H1} in a different way.Indeed, the term with vertex $2 \pom \to 2 \pom$ appears  in QCD when we  calculate the $1/N_c$ corrections, as we have argued in the introduction.  In QCD these corrections generate $2 \pom \to 2 \pom$ of the order of $1/N^2_c$. Hence we come to the hamiltonian of \eq{H1} as  the way to generate the parton cascade in $1/N_c^2$ order of QCD. $\kappa \sim1/N^2_c$ is the natural estimates for this approach.

Bearing this in mind we can conclude that the hamiltonian of \eq{H1}  shows that $1/N^2_c$ correction in QCD can be summed in the framework of the Pomeron calculus.

~


\section{Multiplicity distributions and entropy of produced dipoles.}
  The presentation of the scattering matrix in the form of \eq{POC3} and \eq{POC5}  is very convenient for use of the AGK cutting rules\cite{AGK} for determining the multiplicity distributions of produced dipoles, since each term of the series is the exchange of $k$ Pomerons: 
  \beq \label{POC51}
S\Lb \Y\Rb \,=\intl^\infty_{-\infty}
\frac{d \lambda}{ 2 \sqrt{\pi\,\kappa \,\Y} } \exp\Lb - \frac{ \Lb \kappa  \Y - \lambda\Rb^2}{ 4 \kappa  \Y} \Rb
\sum^\infty_{k=0} \underbrace{(- \kappa)^k \Lb 1\,- \h \kappa (k+2) k (k-1)\Rb
e^{ \Lb 1 - \kappa  - \lambda   \Rb\,k\,\Y}}_{ F_k(\kappa, \Y)};
\eeq

The AGK cutting rules \cite{AGK} allows us to calculate the contributions of $n$-cut Pomerons if we know $F_k$: the contribution of the exchange of $k$-Pomerons to  the cross section. They take the form:
   \begin{subequations} 
    \bea \label{AGKK}
n\,\geq\,1:\sigma^k_n\Lb Y\Rb&=& (-1)^{k-n}\frac{k!}{(n - k)!\,n!}\,2^{k}\, F_k(\gamma, Y)\label{AGKK}\\
n\,=\,0:\sigma^k_0\Lb Y\Rb&=&\Lb -1\Rb^k \Bigg(2^k\,\,-\,\,2\Bigg) F_k(\gamma, Y);\label{AGK0}\\
\sigma_{tot}&=&\,\,2 \sum_{k=0}^\infty (-1)^{k} \,F_k(\gamma, Y);\label{XS}\,
\eea
 \end{subequations} 
where $\sigma_{tot}$ is the total cross section and $\sigma_0$ denotes the cross section with the multiplicity of produced dipoles  which is much less than $\Delta\,Y$. In other words, it is the cross section of the diffraction production.  

The total production of $n$ cut Pomrerons is equal to 
\beq \label{XSAGK}
\sigma^{AGK}_n\,\,=\,\,\sum_{k=n} \sigma^k_n 
\eeq.

The main idea of using the AGK cutting rules is originated from the contribution of the BFKL Pomeon  and s-channel  unitarity for its Gree's function:
\beq \label{UNPOM}
2 \,{\rm Im} G_{\pom}\Lb Y\Rb\,\,=\,\,\sigma_{\pom}\Lb Y\Rb
\eeq
where $\sigma_{\pom}$ is the cross section of the  produced dipoles with the average number$ \bar n = \Delta_{\pom} \,Y$, with $\Delta_{\pom}$ is the intercept of $G_{\pom}$. The multiplicity distribution of these dipoles is the Poisson one, viz.:
\beq \label{PD}
\mathcal{P}_D\Lb n, \Delta_{\pom} Y\Rb\,\,=\,\,\frac{\Lb \Delta_{\pom}\,Y\Rb^n}{n!} e^{ - \Delta_{\pom}\,Y}
\eeq
   
   From \eq{AGKK} and \eq{PD} the multiplicity distribution is equal to 
   \beq \label{MD1}
\sigma_n\,\,=\,\,\sum_{k=1}^{\infty} \sigma^{AGK}_k \Lb Y\Rb \mathcal{P}_D\Lb n, k\, \Delta_{\pom} Y\Rb  \,\,\xrightarrow{ Y \gg 1}  \sigma^{AGK}_{k=n/\Lb \Delta\,Y\Rb} \Lb Y\Rb
\eeq
   
   From \eq{POC51},\eq{AGKK} and \eq{XSAGK} we obtain for $\sigma^{AGK}_n$:
      \bea \label{MD2}
\sigma^{AGK}_n\,\,&=& \,\, \intl^\infty_{-\infty}
\frac{d \lambda}{ 2 \sqrt{\pi\,\kappa \,\Y} } \exp\Lb - \frac{ \Lb \kappa  \Y - \lambda\Rb^2}{ 4 \kappa  \Y} \Rb \Bigg\{ \frac{(2 \zeta  \kappa )^n}{ (2 \zeta  \kappa +1)^{n+1}}\nn\\
& \,\,-\,\,&\h \kappa \frac{\Lb 2\kappa  z\Rb^n }{(2 \kappa  z+1)^{n+4}} \left(8 \kappa ^2 (n+4) z^2-2 \kappa  n (5 n+11) z+(n-1) n (n+2)+16 \kappa ^3 z^3\right)\Bigg\}
\eea
with $\zeta = e^{ \Lb 1 - \kappa \Rb\,\,\Y\,\,-\,\,\lambda}$.

Let us consider for simplicity the first term in $\{\dots\}$. For $ 2 \kappa \zeta \gg 1$ it takes the following form:

 \beq \label{MD3}
\sigma^{AGK}_n\,\,= \,\, \intl^{ (1 - \kappa)\Y + \ln (2 \kappa)}_{-\infty}
\frac{d \lambda}{ 2 \sqrt{\pi\,\kappa \,\Y} } \exp\Lb - \frac{ \Lb \kappa  \Y - \lambda\Rb^2}{ 4 \kappa  \Y} \Rb \frac{1}{2 \zeta  \kappa}\,\exp\Lb - \dfrac{n-1}{ 2\,\kappa\,\zeta} \Rb
\eeq

Integral over $\lambda$ can be taken using the method of steepest descent
with the following equation for the saddle point
\beq
\label{LAMBDASP}
\frac{(n-1) e^{-t-(1-2 \kappa ) Y}}{2 \kappa }-\frac{t}{2 \kappa  Y}-1\,\,=\,\,0
\eeq

This equation can be rewritten as W-Lambert equation:
\beq
\label{LAMBDASP1}
t'_{SP} \,e^{t'_{SP}} \,\,=\,\,\tau\,\, = (n - 1) \,\Y\,e^{ - (1- 4\,\kappa)\Y};~~~~~~t' = t\,+\,2 \kappa \Y\,=\,3\,\kappa\,\Y\,-\,\lambda;
\eeq
with the solution: $t'_{SP}= W \Lb \tau\Rb$ where $W \Lb \tau\Rb$ is the 
W-Lambert function (see section 4.13 of Ref.\cite{OLBC}). This function can be written as the following series:
\beq \label{WF}
W\Lb \tau\Rb\,\,=\,\,\sum^{\infty}_{n=1} \, \tau^n\,\dfrac{\Lb - n\Rb^{n-1}}{n!}
\eeq

     \begin{figure}[ht]
    \centering
  \leavevmode
  \begin{tabular}{c c c}
      \includegraphics[width=5.75cm]{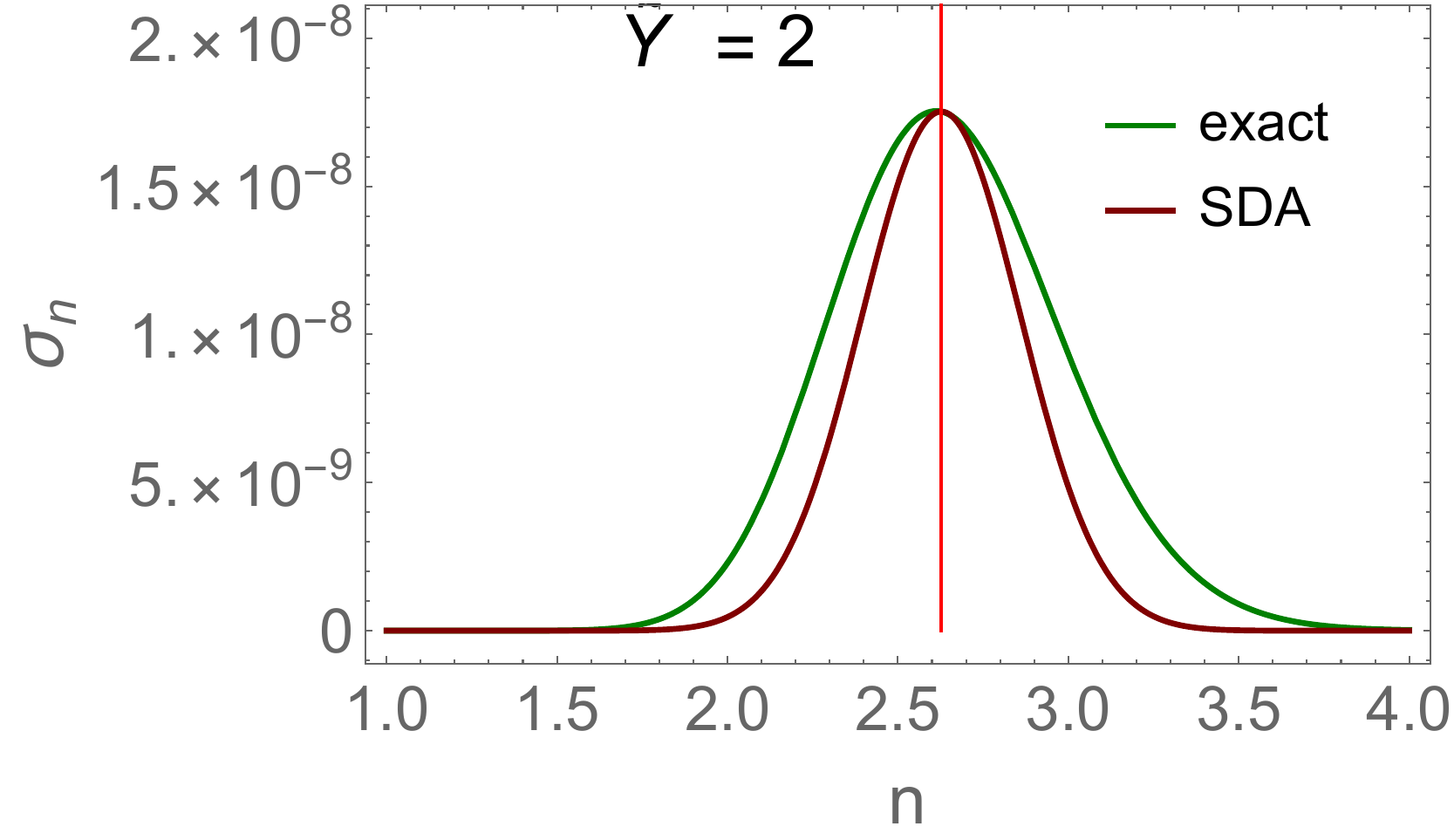} &
       \includegraphics[width=5.5cm]{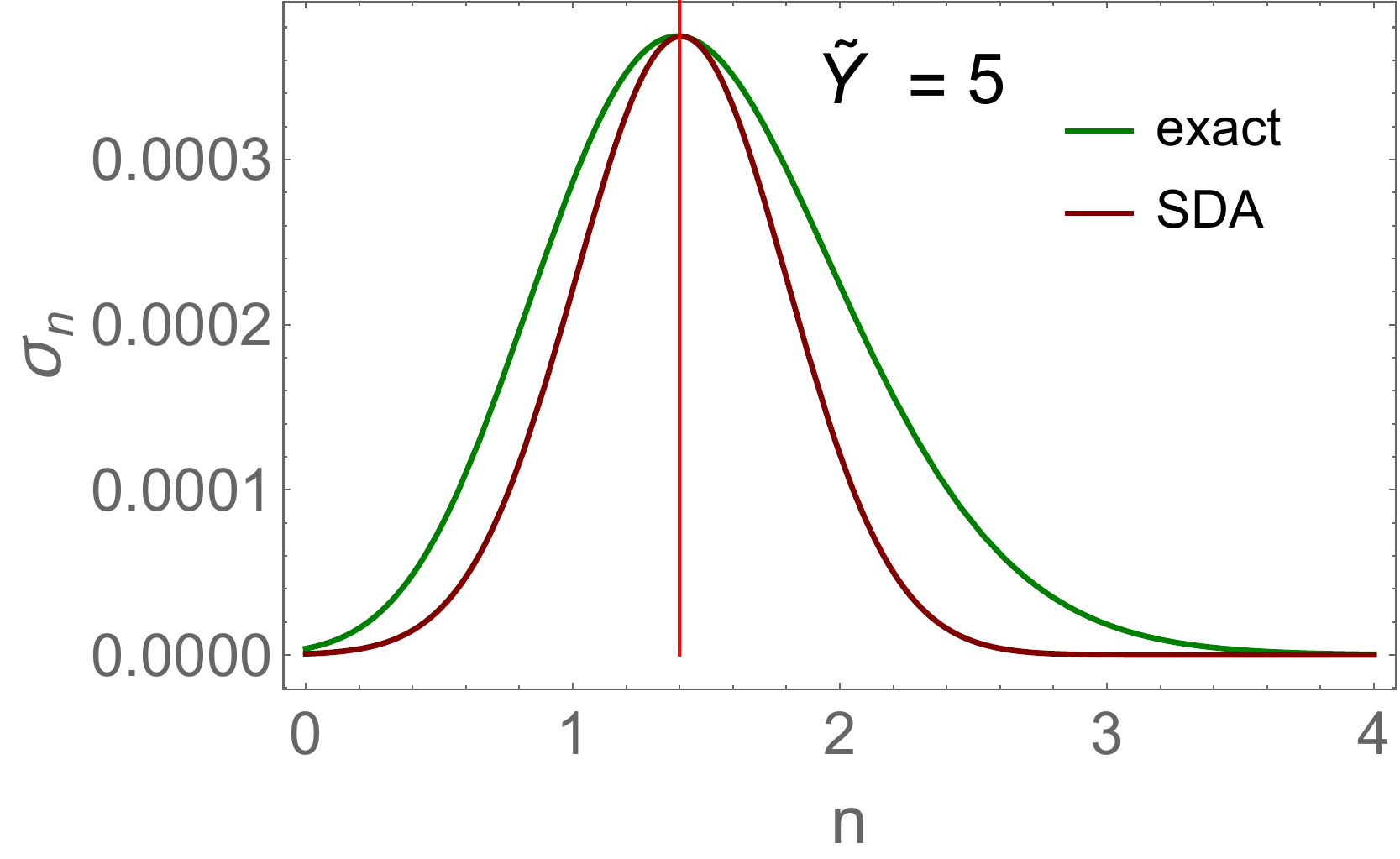}  &
              \includegraphics[width=5.75cm]{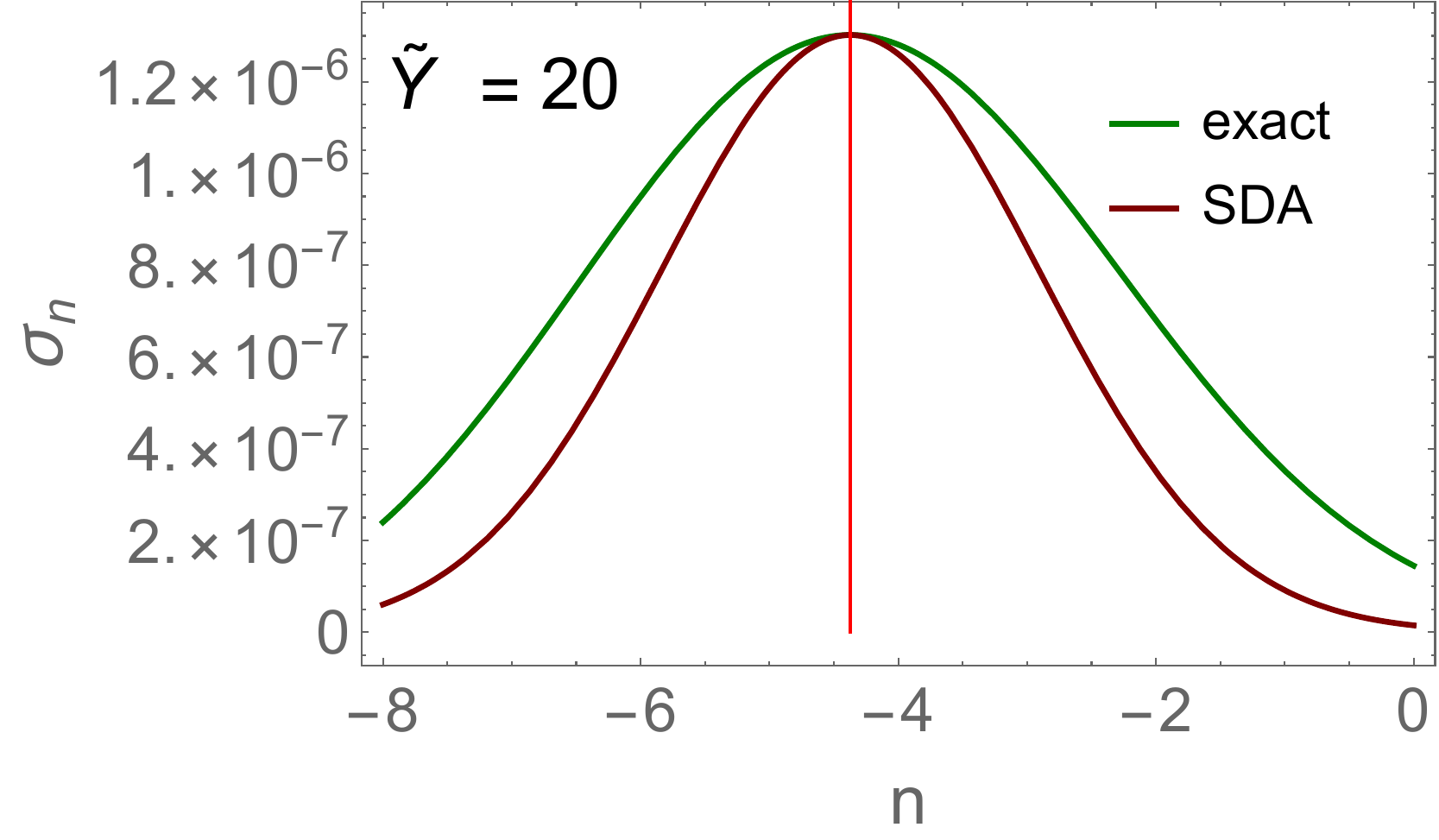} \\
              \fig{xsvst}-a &   \fig{xsvst}-b    &   \fig{xsvst}-c\\
                 \end{tabular}         
                    \caption{ Integrant function of \eq{MD3} versus $t$ at n=100 for different energies. The vertical red line gives the value of $t_{SP}$. }
\label{xsvst}
   \end{figure}

\begin{figure}[ht]
    \centering
  \leavevmode
   \begin{tabular}{c  c}   
      \includegraphics[width=9cm]{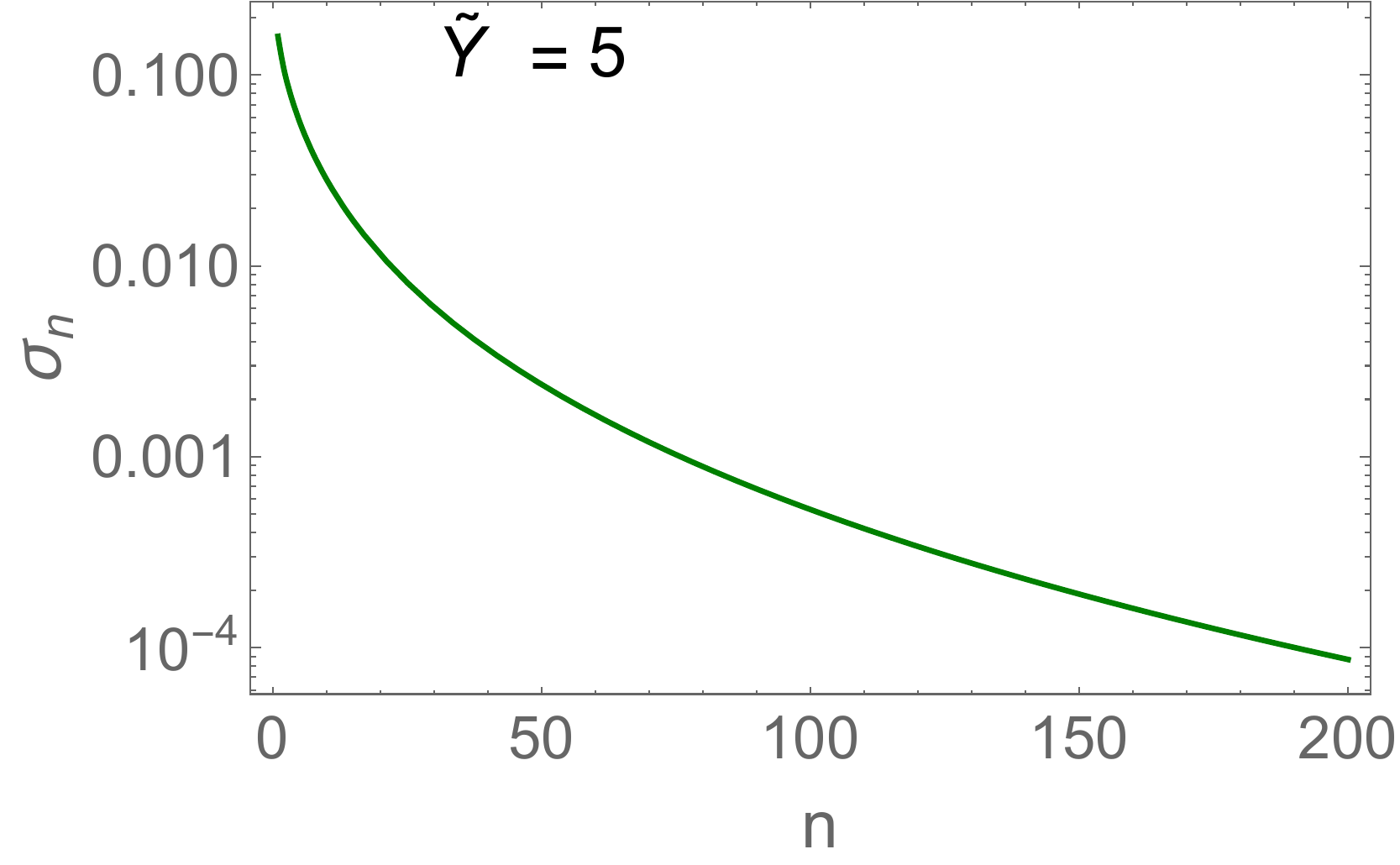} &  \includegraphics[width=8.5cm]{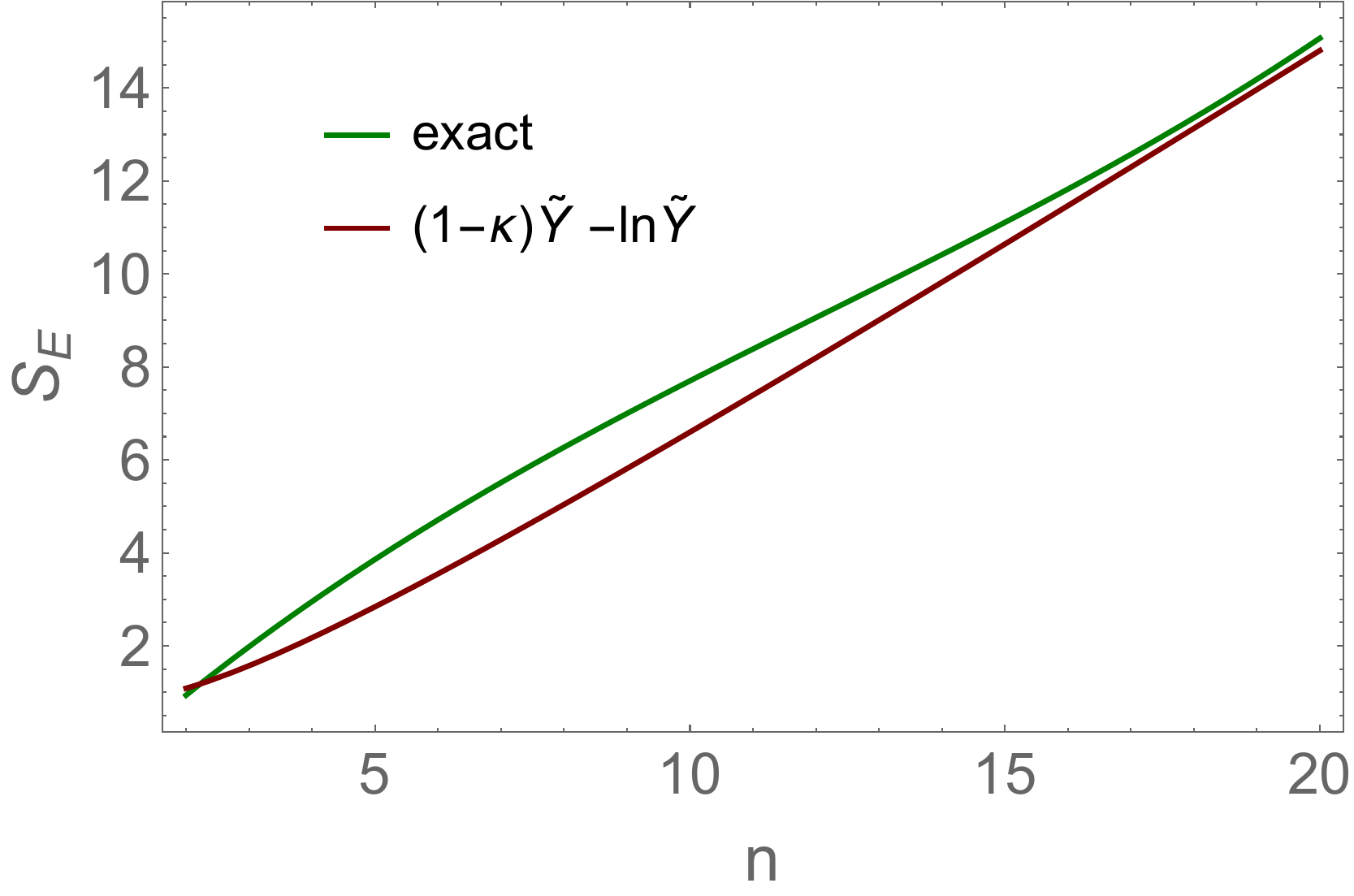} \\
  \fig{xsy5}-a&\fig{xsy5}-b\\        
      \end{tabular}
   \caption{  \fig{xsy5}-a:$\sigma_n$ versus $n$ at $\Y$ =5. \fig{xsy5}-b: The entropy of produced dipoles at different values of $\Y$. }
\label{xsy5}
   \end{figure}

\fig{xsvst} shows that the $t_{SP} $ correctly determine the maximum of the function but the method of  steepest descent leads to more narrow distribution. For small $\tau$ one can see from \eq{WF} that $\sigma_n \propto \exp\Lb - \tau^2 (n-1)^2\Rb$. In \fig{xsy5} we plot $\sigma_n$ for $\Y = 5$. One can see  this kind of the behaviour.

In \fig{xsy5}-b we plot the calculation of the entropy of the produced dipoles:
\beq \label{MD4} 
S_E\,\,=\,\,-\,\sum^{\infty}_{n=1} \ln\Lb  \mathcal{P}_n \Rb \mathcal{P}_n
~~~\mbox{with}~~~ \mathcal{P}_n = \frac{\sigma_n}{\sum^\infty_{n=1} \sigma_n}
\eeq

We compare this calculations with  $S_E = \ln\Lb \zeta(\lambda_{SP}=0)\Rb= (1-\kappa) \Y - \ln \Y $.  This expression is close to the prediction of Ref.\cite{KHLE} but strictly speaking this paper predicts $S_E = \ln \Lb c_1 \Rb= (1+ \kappa)\Y$. Therefore, we can see that the contribution of the enhanced Pomeron diagrams manifest themselves differently in $S_E$ and $c_1$.

~


\section{Conclusions}
The main result of the paper is that we show that the Pomeron calculus can be used to treat the $1/N_C$ corrections in QCD. In the one dimensional model which is a simplification of the QCD approach \cite{LELU1} that gives the description of the high energy interaction both in the framework of the parton cascade (see Refs.\cite{MUDI,MUSA} for details)  and in the Pomeron calculus, we  found the following. First,  the scattering amplitude can be written as the sum of Green's function of $n$-Pomeron exchanges $G_{n \pom} \propto e^{ \omega_n \Y}$ with $\omega_n = \kappa\,n^2$ at $\kappa \ll 1$.  This means that choosing $\kappa = 1/N^2_c$ we can reproduce the intercepts of QCD in $1/N_c$ order.  Second,  the scattering amplitude is an asymptotic series that cannot be sum using Borel approach. We found the general way of summing such series (see \eq{POC5}). Third, we found the set of negative eigenvalues which corresponds to the partonic description of the scattering amplitude. 
Fourth, it turns out that the negative eigenvalues  $\tilde{\omega}_n$ at  $\kappa \ll 1$ are equal to $        
     \tilde{\omega}_n = \omega_{-n} $ (see \eq{RHO15}). Therefore, we can view the expansion of the scattering amplitude in the eigenfunctions with negative eigenvalues as the analytic continuation of the Pomeron calculus to the negative $n$.  Hence, we generalize the analysis of the BFKL and UTM model, given in Ref.\cite{UTMM} , to our approach. Fifth, using AGK \cite{AGK} cutting rules we found the multiplicity distributions of the produced dipoles as well as their entropy $S_E$.  At large $\Y$  $S_E = (1 - \kappa)\,\Y$. This expression is close to the prediction of Ref.\cite{KHLE}: $ S_E = \ln c_1$, where $c_1 $ is the average multiplicity. However,  we show that $\ln c_1 = (1+ \kappa)\,Y$ at large $\Y$. Therefore, we found that the enhanced diagrams of the Pomeron calculus contribute in a different way to $c_1$ and the entropy.
     
    As we have discussed in the introduction both the QCD hamiltonian of \eq{H}-type \cite{LELU1}  as well as the one dimensional incarnation of it  (see \eq{H1})  satisfy the t-channel  but have problems with the s-channel
  unitarities. It is enough to mention that hamiltonian of \eq{H1} is quite different with the ``diamond"    hamiltonian \cite{KLLL2}  which is the best that we know in QCD. However, we do not see how  the correct QCD hamiltonian can change two features of the model: the intercepts $\omega_n \propto (1/N^2_c) n^2$, which is found in QCD and our way of summing the asymptotic series with such kind of intercepts.

     ~

     ~
     
     ~

     ~

     ~
     
        {\bf Acknowledgements}

   We thank our colleagues at Tel Aviv university  for
 encouraging discussions. Special thanks go A. Kovner, M. Li  and M. Lublinsky for stimulating and encouraging discussions on the subject of this paper.   This research    was supported  by 
  Binational Science Foundation  grant \#2022132.

\end{document}